\documentclass[%
reprint,
superscriptaddress,
amsmath,amssymb,
aps,
prb,
]{revtex4-2}

\usepackage{graphicx}% Include figure files
\usepackage{dcolumn}% Align table columns on decimal point
\usepackage{bm}% bold math
\usepackage{hyperref}% add hypertext capabilities
\usepackage{xcolor}
\hypersetup{
  colorlinks,
  citecolor=blue,
  linkcolor=blue,
  urlcolor=magenta}

\usepackage{braket}
\usepackage{textcomp}

\def\markup{0}
\ifnum\markup=1
\newcommand{\red}[1]{\textcolor{red}{#1}}
\usepackage[mathlines]{lineno}% Enable numbering of text and display math
\linenumbers\relax % Commence numbering lines
\else
\newcommand{\red}[1]{#1}
\fi

\begin{document}

\preprint{APS/123-QED}

\title{\red{Coherent all-optical control of the germanium vacancy in diamond}}

\author{C. Adambukulam}
\email[Corresponding Author: ]{cadambukulam@phys.ethz.ch}
\thanks{%
Current Affiliation: Department of Physics, ETH Z\"urich, Otto-Stern-Weg 1, 8093 Z\"urich, Switzerland
}%
\affiliation{%
School of Electrical Engineering and Telecommunications, University of New South Wales, Kensington NSW 2052, Australia
}%

\author{J. A. Scott}
\thanks{%
Current Affiliation: The University of Sydney Nano Institute, The University of Sydney, Camperdown NSW 2006, Australia
}%
\affiliation{%
School of Mathematical and Physical Sciences, University of Technology Sydney, Ultimo NSW 2007, Australia
}%

\author{S. Q. Lim}
\affiliation{%
Centre of Excellence for Quantum Computation and Communication Technology, School of Physics, University of Melbourne, Melbourne VIC 3010, Australia
}%

\author{I. Aharonovich}
\affiliation{%
School of Mathematical and Physical Sciences, University of Technology Sydney, Ultimo NSW 2007, Australia
}%
\affiliation{%
Centre of Excellence for Transformative Meta-Optical Systems (TMOS), Faculty of Science, University of Technology Sydney, Ultimo NSW 2007, Australia
}%

\author{A. Morello}
\affiliation{%
School of Electrical Engineering and Telecommunications, University of New South Wales, Kensington NSW 2052, Australia
}%
 
\author{A. Laucht}
\email{a.laucht@unsw.edu.au}
\affiliation{%
School of Electrical Engineering and Telecommunications, University of New South Wales, Kensington NSW 2052, Australia
}%

\date{\today}

\begin{abstract}
\red{The germanium vacancy in diamond (GeV) is a promising candidate for color center based quantum networking. Yet, like for other group-IV vacancy defects in diamond, achieving fast, high-fidelity qubit operations using traditional magnetic resonance techniques is experimentally challenging due to a weak magnetic dipole and susceptibility to thermally induced decoherence. Here, we perform all-optical control of the GeV and realize Rabi frequencies exceeding $\sim 20$~MHz. We do so by driving the two \texorpdfstring{$\Lambda$}{Lambda}-systems of the GeV, simultaneously and apply this to probe the spin coherence ($T_2^*=224\pm14$~ns, $T_2^{\rm H}=11.9\pm0.3$~\textmu s). Our control scheme is applicable to other color centers and particularly, other group-IV defects for which, the scheme may be optimized to improve all-optical control in these systems.}
\end{abstract}

%\keywords{Suggested keywords}%Use showkeys class option if keyword
                              %display desired
\maketitle

\section{Introduction}

Realizing a quantum network will require developing light-matter interfaces consisting of coherent matter-qubits capable of emitting photons to perform long distance entangling gates~\cite{Kimble2008, Duan2001, Barrett2005}. Solid-state systems are particularly desirable in this role, as they are integrable into various micro- and nanostructures~\cite{aharonovich2016solid}, while systems that operate in the optical domain cover larger distances than their microwave counterparts owing to low-loss fiber optics~\cite{Lambert2020}. As optically-active solid-state systems with outstanding spin coherence times~\cite{Abobeih2018, Nagy2018}, color centers are well suited to deployment as light-matter interfaces. Of the various color centers, group-IV split vacancy defects in diamond have attracted recent attention due to their excellent spin~\cite{Sukachev2017} and spectral properties~\cite{Chen2022}. These defects possess an inversion symmetry derived insensitivity to electric fields that enables long coherence times~\cite{Sukachev2017} and lifetime-limited spectral purity~\cite{Bradac2019}, even when incorporated into an environment with significant electrical noise such as a nanophotonic cavity~\cite{Trusheim2020, Bhaskar2017, Zhang2017} -- a key step in the efficient generation of long-distance entanglement~\cite{Kimble2008}. Additionally, short optical lifetimes on the order of nanoseconds and a high zero-phonon line emission (Debye-Waller factors exceeding $\sim 70\%$~\cite{Neu2011}) make group-IV defects promising candidates for high-efficiency spin-photon interfaces in a quantum network~\cite{knaut2023} and as a platform for microwave-to-optical transduction~\cite{Neuman2021}.

\begin{figure}[!ht]
\centering
\includegraphics{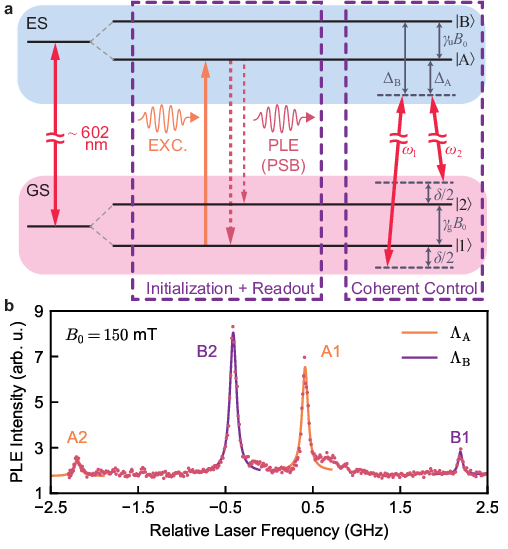}
\caption{\label{fig::Initialization} 
(a) Level structure of the GeV with the initialization and readout, and the optical control schemes shown. In the initialization and readout scheme a resonant laser (EXC.) cycles the A1 transition with the resulting phonon side band (PSB) emission (PLE) collected. For optical control, two equal power (a consequence of the experimental setup~\cite{SuppMat, Adambukulam2023}) lasers with frequencies $\omega_1$ and $\omega_2$ are applied to drive the Raman transitions of $\Lambda_{\rm A}$ and $\Lambda_{\rm B}$, simultaneously.
(b) PLE spectrum of the GeV used in this work.}
\end{figure}

In this article, we focus on a single negatively-charged germanium vacancy (GeV). As a group-IV center with a ground-state spin-orbit coupling of $\sim 165$~GHz~\cite{Maity2018}, the GeV must be operated at temperatures $\ll 300$~mK; in a dilution refrigerator, to \textit{freeze-out} thermal excitation into a higher-energy orbital that would otherwise introduce decoherence~\cite{Becker2018}. At mK, cooling power is limited and fast electron spin operations may not be compatible with traditional magnetic resonance techniques due to power dissipation within the cryostat~\cite{Vallabhapurapu2021}. \red{In particular, the heat produced from applying microwaves to control the spin would likely introduce decoherence by thermally populating the higher energy states~\cite{Nguyen2019}.} Weakly strained group-IV centers are particularly affected due to exhibiting an almost non-existent magnetic coupling between the ground-state spin sub-levels~\cite{Rosenthal2023, Meesala2018} \red{while having the lowest possible energy separation between the spin-1/2 ground-state and the higher energy orbital state~\cite{Hepp2014thesis}}. Yet, fast spin rotations are necessary for robust quantum information processing. Fortunately, optical control techniques offer a pathway to realizing these fast rotations by leveraging the orders of magnitude larger electric dipole moment of the optical transition over the spin magnetic moment. \red{Such control techniques have found use in a wide variety of optically active systems including trapped ions~\cite{Ballance2016} and neutral atoms~\cite{Jones2007}, self-assembled quantum dots~\cite{Press2008} and other color centers such as the nitrogen~\cite{Chu2015, Yale2013}, silicon~\cite{Becker2018} and tin vacancy~\cite{Debroux2021} defects \red{in diamond}. Here, we extend the use of these experimental techniques to the GeV by demonstrating all-optical initialization, readout and coherent control of the GeV. In doing so, we achieve Rabi frequencies exceeding $\sim 20$~MHz, with which we probe the coherence of the GeV in an off-axis magnetic field and at $\sim 25$~mK.}

\begin{figure}[!h]
\centering
\includegraphics{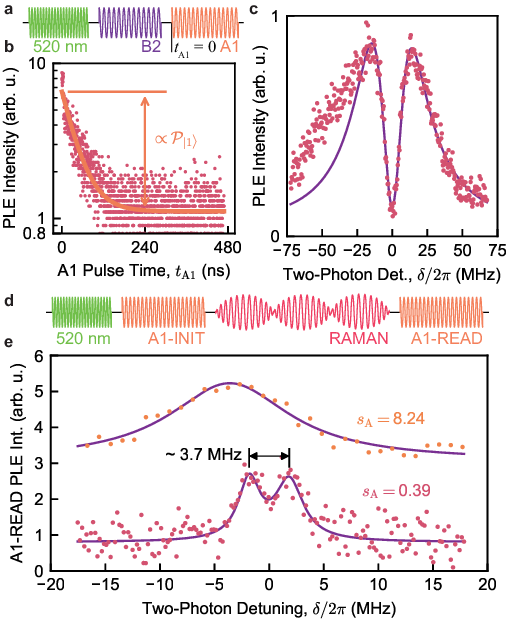}
\caption{\label{fig::RamanSpectrum} 
(a) Pulse sequence used to measure the spin initialization into $\ket{2}$ plotted in (b). Here, and throughout this article, the $\sim 520$~nm laser pulse stabilizes the GeV charge configuration~\cite{Chen2019}.
(b) A histogram of the photon arrival times, measured during the A1 spin-pumping pulse that transfers the population into $\ket{2}$. Here, $\mathcal{P}_{\ket{1}}$ refers to the population in $\ket{1}$ at $t_{\rm A1} = 0$~ns.
(c) CPT spectrum performed with $\Lambda_{\rm A}$. The data (red dots) is fit (purple curve) to the steady state of the CPT Lindblad superoperator~\cite{SuppMat}. 
(d) Pulse sequence used to measure the Raman spectrum presented in (e). 
(e) Raman spectra measured with a Raman power of $s_{\rm A} = 0.39$ (red dots) and $s_{\rm A} = 8.24$ (orange dots). The one-photon detunings were set to \red{$\Delta_{\rm A}/2\pi \sim -790$~MHz and $\Delta_{\rm B}/2\pi \sim 970$~MHz.}}
\end{figure}

\section{Initialization and Readout}

The GeV may be viewed as consisting of a ground- and an excited-state spin-1/2 system [see Fig.~\ref{fig::Initialization}(a)]~\cite{Adambukulam2023, Hepp2014, Chen2022, Parker2023}. Following the standard nomenclature, we label the ground-state, spin-down eigenstate as $\ket{1}$ and its excited-state counterpart $\ket{\rm A}$, and likewise for the spin-up eigenstates, $\ket{2}$ and $\ket{\rm B}$. In a static magnetic field, $B_0$, the Zeeman effect lifts the ground-state spin degeneracy by $\gamma_{\rm{g}} B_0$ and its excited-state equivalent by $\gamma_{\rm{u}} B_0$. Here, $\gamma_{\rm{g,u}} = \sqrt{\gamma_{\parallel,\rm{g,u}}^2 \cos^2{\theta}  + \gamma_{\perp,\rm{g,u}}^2 \sin^2{\theta}}$, where $\gamma_{\perp}$ and $\gamma_{\parallel}$ are the transverse and axial components of the g-tensor in the ground-state (excited-state) and $\theta$ is the angle between the magnetic field and the GeV high-symmetry axis -- in this case, $\theta \sim 70.5^\circ$ while $B_0 = 150$~mT. Owing to the strain dependence of $\gamma_\perp$, the Zeeman splitting and spin quantization axis differ between the ground- and excited-state, giving rise to two spin-\textit{conserving} optical transitions that correspond to $\ket{1} \leftrightarrow \ket{\rm A}$ (A1) and $\ket{2} \leftrightarrow \ket{\rm B}$ (B2), and two weakly allowed spin-\textit{flipping} transitions $\ket{2} \leftrightarrow \ket{\rm A}$ (A2) and $\ket{1} \leftrightarrow \ket{\rm B}$ (B1). All four transitions are spectrally resolvable,, as seen in Fig.~\ref{fig::Initialization}(b).

For spin initialization and readout, we resonantly drive A1 as shown in Fig.~\ref{fig::Initialization}(a) and ~\ref{fig::RamanSpectrum}(a). This pumps any population from $\ket{1}$ into $\ket{2}$ via optical relaxation through A2. As relaxation by A1 is favored, the transition will cycle and emit several photons before relaxation into $\ket{2}$ quenches the emission. We observe this in Fig.~\ref{fig::RamanSpectrum}(b), where we plot a histogram of the photoluminescence excitation (PLE) intensity during an A1 pulse and find that it decays exponentially with a decay time of $\sim 42$~ns~\footnote{Before applying the A1 pulse, the system was prepared in the $\ket{1}$ state using a B2 pulse, as indicated in Fig.~\ref{fig::RamanSpectrum}(a).}. The PLE intensity at $t_{\rm A1} = 0$~ns is proportional to the population of $\ket{1}$ prior to the A1 pulse, $\mathcal{P}_{\ket{\rm 1}}$, and therefore, by measuring this amplitude (referred to by ``A1-READ PLE Int.'' in later figures), we perform spin readout. By driving A1 until the emission is quenched, we have initialized the spin into $\ket{2}$.

To optically access the ground-state spin sub-levels, we first note that A1 and A2 as well as B2 and B1 form two $\Lambda$-systems that we label $\Lambda_{\rm{A}}$ and $\Lambda_{\rm{B}}$. When driving both transitions of a $\Lambda$-system with two equal-powered lasers, we induce a transition between the ground-state spin sub-levels -- here, referred to as the Raman drive -- as shown in Fig.~\ref{fig::Initialization}(a). In this particular system, the presence of two $\Lambda$-systems will result in the ground-state spin sub-levels being driven via both $\Lambda$-systems simultaneously. For the two Raman lasers with frequencies, $\omega_1$ and $\omega_2$ and the four optical transitions with frequencies, $\omega_{\rm{A1}}$, $\omega_{\rm{A2}}$, $\omega_{\rm{B2}}$ and $\omega_{\rm{B1}}$,  \red{we define the one-photon detunings of the two $\Lambda$-systems as $2\Delta_{\rm{A}} = (\omega_{\rm{A1}} + \omega_{\rm{A2}}) - (\omega_1 + \omega_2)$ and $2\Delta_{\rm{B}} = (\omega_{\rm{B1}} + \omega_{\rm{B2}}) - (\omega_1 + \omega_2)$, and the two-photon detuning as $\delta = \gamma_{\rm g} B_0 -(\omega_1 - \omega_2)$~\cite{SuppMat}}. These parameters hence define the system Hamiltonian in the rotating frame as 
\begin{equation}
\begin{split}
    \frac{H}{\hbar} &= \frac{\delta}{2} \sigma_z + \Delta_{\rm{A}} \ket{\rm A}\!\!\bra{\rm A} + \Delta_{\rm{B}} \ket{\rm B}\!\!\bra{\rm B}  \\
    &+ \frac{1}{2}\big(\Omega_{\rm{A1}} \ket{1}\!\!\bra{\rm A} + \Omega_{\rm{A2}} \ket{2}\!\!\bra{\rm A} + \rm{h.c.} \big)  \\
    &+ \frac{1}{2}\big(\Omega_{\rm{B1}} \ket{1}\!\!\bra{\rm B} + \Omega_{\rm{B2}} \ket{2}\!\!\bra{\rm B} + \rm{h.c.} \big),
\end{split}
\label{eq::Hamiltonian}
\end{equation} where $\Omega_{\rm{A1}},...,\Omega_{\rm{B2}}$ are the optical Rabi frequencies and $\sigma_z = \ket{2}\!\!\bra{2} - \ket{1}\!\!\bra{1}$. It should be noted that \red{$\Delta_{\rm{A}} = \Delta_{\rm{B}} -  \gamma_{\rm{u}} B_0$}.

\section{Two-Laser Spectroscopy and Coherent Control}

We begin by performing coherent population trapping (CPT)~\cite{pingault2014all, Xu2008} on $\Lambda_{\rm A}$ to locate the Raman transition. To do so, we set $\Delta_{\rm{A}} = 0$~MHz, sweep $\delta$ -- by sweeping $\omega_1$ and $\omega_2$, simultaneously -- and record the steady state PLE intensity in a continuous wave measurement. The results of which are presented in Fig.~\ref{fig::RamanSpectrum}(c). We observe diminished emission in the vicinity of $\delta = 0$~MHz in the form of a \textit{dip} in the spectrum that is consistent with the formation of the CPT dark state. \red{The deviation of the data (Fig.~\ref{fig::RamanSpectrum}(c), red dots) from the fit (Fig.~\ref{fig::RamanSpectrum}(c), purple line) is due to the excitation power fluctuating as $\delta$ is swept.}

\begin{figure}[!bt]
\centering
\includegraphics{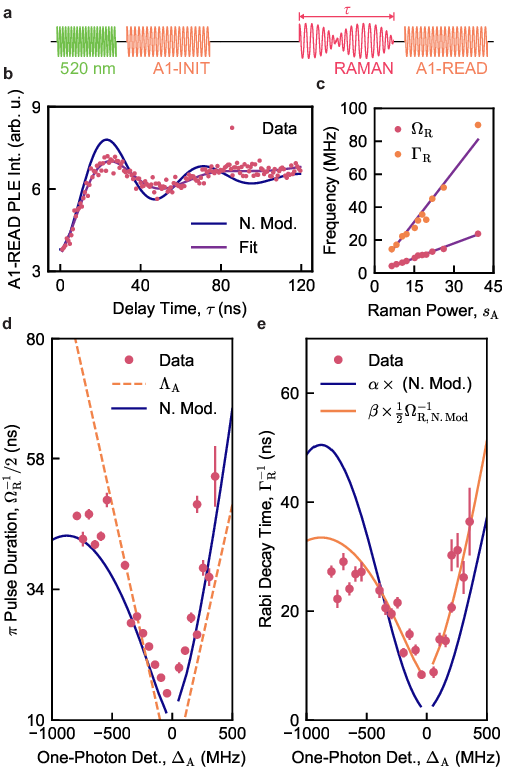}
\caption{\label{fig::Rabi} 
(a) Pulse sequence used to measure Raman-Rabi oscillations. 
(b) Rabi oscillation measured with \red{$\Delta_{\rm A}/2\pi \sim -240$~MHz and $\Delta_{\rm B}/2\pi \sim 1520$~MHz} and with a Raman power of $s_{\rm A} \sim 17$. The data (red dots) is fit to a decaying sinusoid (purple curve). \red{Also plotted is the simulated and scaled population of $\ket{1}$ during the Raman drive (N. Mod, blue line). The simulation is performed as described in Sec. V of Ref.~\cite{SuppMat}}.
(c) Rabi frequency ($\Omega_{\rm R}$, red dots) and Rabi decay rate ($\Gamma_{\rm R}$, orange dots) as measured for various Raman drive powers with the corresponding linear fits (purple lines). The one-phonon detunings are the same as in (b).
(d) The experimentally measured dependence of the $\pi$-pulse duration (red dots) on the one-photon detuning plotted alongside the values predicted by \red{the same simulation model as in (b) (N.Mod, solid blue line)} and the values computed when only $\Lambda_{\rm A}$ (the first term in Eq.~\ref{eq::rabi_frequency_2}) is present (dashed orange line). 
(e) The measured (red dots) one-photon detuning dependence of the Rabi decay time in addition to the \red{simulated decay time (N. Mod, blue line, scaled by $\alpha = 1/3$)} and the \red{simulated $\pi$-pulse duration calculated from (c)} (orange dashed line, scaled by $\beta \sim 0.76$). The data in (d) and (e) were measured with a Raman drive power of $s_{\rm A} \sim 18$.}
\end{figure}

Prior to entering the dark state, coherent oscillations between $\ket{1}$ and $\ket{2}$ should occur. However, these decay very quickly as they are limited by the nanosecond optical decay time of the excited state and cannot be observed under continuous excitation. These Raman-Rabi oscillations continue to occur even when non-zero one-photon detunings are used to reduce excitation into the excited state. Hence, we add a one-photon detuning and measure the resulting Raman spectrum using the pulse sequence in Fig.~\ref{fig::RamanSpectrum}(d). Here, the spin is first initialized into $\ket{2}$, then coherently driven with a fixed duration Raman pulse of varying $\delta$ and subsequently measured. The spectrum for one-photon detunings of \red{$\Delta_{\rm A}/2\pi \sim -790$~MHz and $\Delta_{\rm B}/2\pi \sim 970$~MHz} is plotted in Fig.~\ref{fig::RamanSpectrum}(e), where we observe two peaks centered at $\delta = 0$~MHz for low excitation powers (red dots). As the Raman spectrum provides greater resolution than CPT~\cite{Adambukulam2023, Agapev1993}, we are able to detect the hyperfine structure of a strongly coupled \textsuperscript{13}C nuclear spin~\cite{karim2023}, which introduces a $\pm 1.9$~MHz nuclear state dependent shift to the Raman resonance~\cite{Adambukulam2023} for the two \textsuperscript{13}C spin eigenstates. In addition to power broadening, higher Raman powers introduce a so-called AC Stark shift to the Raman resonance frequency [see Fig.~\ref{fig::RamanSpectrum}(\red{e}) orange dots] that is a consequence of $\Omega_{\rm A1} \neq \Omega_{\rm A2}$ and $\Omega_{\rm B2} \neq \Omega_{\rm B1}$.

We now directly measure Raman-Rabi oscillations by varying the Raman pulse duration while fixing $\delta$ to be on resonance with the Raman transition [see Fig.~\ref{fig::Rabi}(a)]. As shown in Fig.~\ref{fig::Rabi}(b), the spin population in $\ket{1}$ oscillates as the Raman-pulse duration is varied. The linear dependence of the Raman-Rabi frequency, $\Omega_{\rm R}$, with respect to the Raman power [see Fig.~\ref{fig::Rabi}(c)] confirms that these are Raman-Rabi oscillations. Also plotted in Fig.~\ref{fig::Rabi}(c) is the Rabi decay rate, $\Gamma_{\rm R}$, which is extracted from the decay of the oscillations and is proportional to Raman power. Therefore, this decay is induced by the drive and \red{ultimately places a limit on the control fidelity.}

Due to the specific strain configuration of our GeV, the system exhibits $\omega_{\rm{A2}} < \omega_{\rm{B2}} < \omega_{\rm{A1}}$ [see Fig.~\ref{fig::Initialization}(a)]. Hence, $\Lambda_{\rm{A}}$ and $\Lambda_{\rm{B}}$ cannot be separately driven. For a single $\Lambda$-system with excitation power, $s$ (expressed in terms of the saturation power of the constituent transitions of the $\Lambda$-system as a quantity known as the saturation parameter), natural linewidth, $\Gamma$, and one-photon detuning, $\Delta$, the Raman-Rabi frequency is $\frac{s \Gamma^2}{4\Delta}$ for $|\Delta| \gg \sqrt{s}\Gamma$~\cite{SuppMat}. In a multi-$\Lambda$-system, the Raman-Rabi frequency is the sum of the Rabi frequencies over all $\Lambda$-systems~\cite{Chu2015, Press2008}. In this case, the Raman-Rabi frequency (see Ref.~\onlinecite{SuppMat} for derivation) is
\begin{equation}
    \Omega_{\rm R} 
    \approx \frac{s_{\rm A} \Gamma_{{\rm eff},\rm A}^2}{4 \Delta_{\rm A}} \frac{1}{\sqrt{\eta_{\rm A}}} + 
    e^{i\varphi} \frac{s_{\rm B} \Gamma_{{\rm eff},\rm B}^2}{4 \Delta_{\rm B}} \frac{1}{\sqrt{\eta_{\rm B}}}.
    \label{eq::rabi_frequency_2}
\end{equation}
Here, $s_i$, $\Delta_i$, $\eta_i$ and $\Gamma_{{\rm eff},i}$ are the saturation parameter, one-photon detuning, branching ratio and effective linewidth belonging to $\Lambda_{i}$. For the GeV in this \red{work}, $s_{\rm A} \sim s_{\rm B}$ and likewise, $\eta_{\rm A} \sim \eta_{\rm B}$~\cite{SuppMat}. The term, $e^{i\varphi}$, is added to account for the phase difference between constituent transitions of the two $\Lambda$-systems.

In Fig.~\ref{fig::Rabi}(d), we plot the measured Rabi $\pi$-times for various one-photon detunings and observe a quadratic dependence with respect to $\Delta_{\rm A}$. \red{By extracting the $\pi$-time from a numerical simulation of the time evolution of Eq.~\ref{eq::Hamiltonian} [as described in Ref.~\cite{SuppMat}; see also Fig.~\ref{fig::Rabi}(b), blue solid line], we are able to recreate the observed dependence by assuming $\varphi = \pi$. In other words, $\Lambda_{\rm A}$ and $\Lambda_{\rm B}$ possess opposite phases. This effect can occur when a GeV is operated in our experimental conditions~\cite{SuppMat} and it should be noted that a similar effect has been observed in nitrogen vacancy centers~\cite{Chu2015}. We plot the simulated results along side the experimental data in Fig.~\ref{fig::Rabi}(d) (blue solid line) and note that it largely agrees with the trend of the experimentally measured $\pi$-times. A level of disagreement between the simulated and experimental results is expected given that the modeling assumes that $\Omega_{\rm A2(B1)} = \Omega_{\rm A1(B2}/\sqrt{\eta_{\rm A(B)}}$. This does not account for the overlap of the laser polarization and the optical dipole of each transition. Additionally, fluctuations in excitation power as the measurement is carried out, as well as errors in $\delta$, may further the disagreement. Nonetheless, the simulation accounts for the experimentally observed trend better than a model containing only a single $\Lambda$-system [Fig.~\ref{fig::Rabi}(d), dashed orange line] or the double $\Lambda$-system with $\varphi = 0$; where the $\pi$-time would approach $\infty$ as $\Delta_{\rm A} \rightarrow \gamma_{\rm u} B_{0}/2 \sim -880$~MHz.}

We plot the corresponding Rabi decay times in Fig.~\ref{fig::Rabi}(e) and observe a quadratic dependence on $\Delta_{\rm A}$. This is in contrast to the \red{decay extracted from numerical simulations [Fig.~\ref{fig::Rabi}(e), solid blue line] which predicts a decay that is three times slower and has a different relationship to $\Delta_{\rm A}$. This disagreement between the simulated Raman-Rabi decay and its experimental counterpart can also be seen in Fig.~\ref{fig::Rabi}(b) where the experimentally measured Rabi decays significantly faster than in the simulation. An analytical treatment of the Raman-Rabi decay performed in Ref.~\cite{SuppMat} indicates that were off-resonant scattering~\cite{Foot2005} the dominant decoherence mechanism, the Rabi decay rate should follow}
\begin{equation}
    \Gamma_{\rm R} = \frac{\kappa_{\rm A}}{\Delta_{\rm A}^2} + \frac{\kappa_{\rm B}}{\Delta_{\rm B}^2}.
    \label{eq::raman_decay_prop_const}
\end{equation}
\red{Here, $\kappa_{\rm A(B)}$ are real and positive proportionality constants. The simulated Rabi decay times follow the trend described by Eq.~\ref{eq::raman_decay_prop_const}, however, the measured values are instead proportional to the Raman-Rabi frequency. This can be seen in Fig.~\ref{fig::Rabi}(e) (solid orange line). Therefore, we rule out off-resonant scattering as the source of the Rabi decay and in Ref.~\cite{SuppMat}, we similarly rule out cross-talk. The proportionality between the Rabi decay time and the $\pi$-time suggests that Raman excitation noise is the cause of the decay. Given the means by which we generate the two Raman lasers~\cite{SuppMat}, relative phase and frequency jitter between the two lasers are negligible while correlated frequency fluctuations are of limited consequence~\cite{Dalton1982}.  However, given that the Raman-Rabi frequency is directly proportional to the Raman drive power (see Eq.~\ref{eq::rabi_frequency_2}), we expect that Raman excitation intensity noise} -- likely from cryostat-induced vibration of the sample relative to the focal spot -- is the dominant Rabi decay mechanism.

\begin{figure}[t]
\centering
\includegraphics{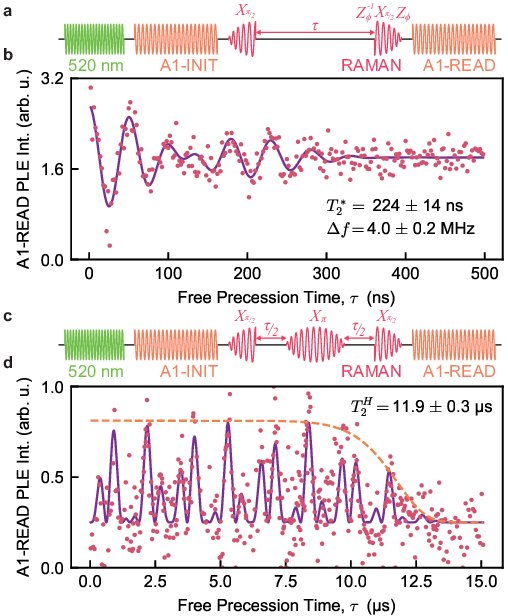}
\caption{\label{fig::Coherence} 
Pulse sequences used to measure the free induction decay (a) and Hahn echo (c) coherence times. Here, $X_{\theta}$ and $Z_{\theta}$ refer to a qubit rotation of angle $\theta$ around the $x$- or $z$-axes of the Bloch sphere. 
(b) Free induction decay. The data (red dots) is fit (purple line) to the sum of two decaying sinusoids with $\Delta f$ denoting the difference in the two sinusoid frequencies. A linear $\tau$-dependent phase shift of $\phi/\tau = 5$~MHz is applied to the second $X_{\pi/2}$ pulse. The one-photon detuning is \red{$\Delta_{\rm A}/2\pi \sim -790$~MHz ($\Delta_{\rm B}/2\pi \sim 970$~MHz}) while the Raman power was $s_{\rm A} \sim 49$.
(d) Hahn echo. The data (red dots) is fit to an appropriate model (purple line) with the decay envelope (orange dashed line) shown. Here, \red{$\Delta_{\rm A}/2\pi \sim -790$~MHz and $\Delta_{\rm B}/2\pi \sim 970$~MHz} while $s_{\rm A} \sim 18$.}
\end{figure}

\section{Spin Coherence}

Finally, we investigate the spin coherence of the GeV. We begin by using the pulse sequence in Fig.~\ref{fig::Coherence}(a) to measure the free induction decay [see Fig.~\ref{fig::Coherence}(b)], where the free precession time, $\tau$, is increased. We observe a beating oscillation in the free induction decay (FID) with frequencies $f_{\rm{FID}}^{(\pm)} = f_{\phi} + f_{\rm{Stark}} \pm \Delta f/2$. Here, $f_{\phi} = 5$~MHz is the $\tau$-dependent phase shift applied to the second $\pi/2$-pulse, $f_{\rm{Stark}} \sim 14$~MHz is the AC Stark shift during the Raman drive, and $\Delta f$ is the hyperfine splitting from the aforementioned \textsuperscript{13}C spin. The Stark shift is only present during the $\pi/2$-pulses which are tuned to the Stark shifted resonance to ensure the proper rotation angle is achieved [comp. Fig.~\ref{fig::RamanSpectrum}(e)]. Consequently, the qubit is detuned by the missing Stark shift during free precession. By fitting the FID to the sum of two exponentially decaying sinusoids, we extract $T_2^* = 224 \pm 14$~ns. This value is typical for group-IV defects in a largely misaligned magnetic field~\cite{Sukachev2017, Debroux2021}. We add a refocusing $\pi$-pulse to the sequence [see Fig.~\ref{fig::Coherence}(c)] to measure the Hahn echo coherence time and plot the results in Fig.~\ref{fig::Coherence}(d). We observe coherence revivals that correspond to \red{the} anisotropic hyperfine coupling to a proximal \textsuperscript{13}C nucleus. The Hahn echo signal is proportional to $e^{-(\tau/T_2^{\rm H})^\beta}  \sin^2{(\kappa_{-} \tau)} \sin^2{(\kappa_{+} \tau)}$ where, $\kappa_{\pm}$ are the \textsuperscript{13}C nuclear magnetic resonance frequencies~\cite{Wood2022, Rowan1965}. From the fit, we extract $T_2^{\rm H} = 11.9 \pm 0.3$~{\textmu}s, as expected for a group-IV defect in a largely off-axis magnetic field~\cite{Debroux2021}.

\section{Conclusion}

Many optically active spin-1/2 systems can form a double $\Lambda$-system~\cite{Lagoudakis2017, Xu2013} -- most notably, the group-IV split vacancy defects where it is common for $\gamma_{\rm u} < \gamma_{\rm g}$. \red{In the case of the group-IV defects, much of the work has focused on operating regimes that reduce these systems to single $\Lambda$-systems~\cite{Debroux2021, Becker2018, Becker2016}. However, outside of such regimes, full consideration of both $\Lambda$-systems is required to successfully apply all-optical control techniques to these systems. Given that typically $\Lambda_{\rm A}$ and $\Lambda_{\rm B}$ are almost identical in terms of the magnitude of optical interaction strengths (i.e. $|\Omega_{\rm A1(A2)}| = |\Omega_{\rm B2(B1)}|$ and $\Gamma_{\rm A1(A2)} = \Gamma_{\rm B2(B1)}$ where $\Gamma_i$ represents the natural linewidth of the transition, $i$), in the case of $\varphi = \pi$, Raman-Rabi oscillations rapidly decay by way of off-resonant scattering in the large $\Delta_{\rm A(B)}$ limit (consider the ratio of Eq.~\ref{eq::rabi_frequency_2} and Eq.~\ref{eq::raman_decay_prop_const} when $s_{\rm A}\Gamma_{\rm eff, A}^2 = s_{\rm B}\Gamma_{\rm eff, B}^2$ and $\kappa_{\rm A} = \kappa_{\rm B}$). Fortunately, the relative phases of the two $\Lambda$-systems are strain and magnetic field dependent~\cite{SuppMat}, and therefore, one can instead tune the strain~\cite{Meesala2018} or magnetic field such that $\varphi \sim 0$. In this case, the rate of off-resonant scattering relative to the Raman-Rabi frequency is proportional to $1/\Delta_{\rm A(B)}$. Consequently, $\Delta_{\rm A(B)}$ may be chosen to be sufficiently large so that dephasing, rather than off-resonant scattering, is the dominant decoherence mechanism. Here, the double $\Lambda$-system would prove advantageous by offering a two-fold increase in the achievable Raman-Rabi frequency, allowing for higher fidelity gate operations. In a similar vein, we observe Raman-Rabi frequencies that are almost twice as fast as would have been produced by single $\Lambda$-system in the regime where $\Delta_{\rm A} \sim -\Delta_{\rm B}$ [see Fig.~\ref{fig::Rabi}(d)] -- a consequence of $\varphi = \pi$.}

\red{Ultimately, we have demonstrated coherent all-optical control of the electronic spin of the GeV and used it to probe the coherence of the system. In doing so, we have achieved Raman-Rabi oscillations exceeding $\sim 20$~MHz, faster than would be possible in a single $\Lambda$-system without requiring higher Raman drive powers. More importantly, we achieve these quick Rabi oscillations with drive powers on the order a few \textmu W. This is significantly lower than the power required for microwave control of the GeV~\cite{Senkalla2024}. The ability to achieve fast Rabi oscillations at low drive powers is significant} as it enables fast, all-optical quantum gates -- a requirement for robust quantum information processing -- in the low cooling power environment of a dilution refrigerator and without causing localized heating that would introduce further decoherence. While we recognize that currently the GeV exhibits significant Raman induced decoherence, we note that this is not intrinsic to the physical system and improvements to the experimental setup would minimize this effect. Moreover, as the optical response of the GeV is strain and magnetic field dependent, the Raman transitions may be optimized to significantly improve the application of optical control techniques to the GeV or more broadly, the group-IV defects.

\begin{acknowledgments}
This work was funded by the Australian Research Council (grant no. CE170100012, CE200100010, FT220100053) and the Office of Naval Research Global (grant no. N62909-22-1-2028). We acknowledge the facilities, and the scientific and technical assistance provided by the NSW node of the Australian National Fabrication Facility (ANFF), and the Heavy Ion Accelerators (HIA) nodes at the Australian National University. ANFF and HIA are supported by the Australian Government through the National Collaborative Research Infrastructure Strategy (NCRIS) program. C. A. and A. L. acknowledge support from the University of New South Wales Scientia program.
\end{acknowledgments}

\bibliography{references}

\end{document}